# Novel Cu-based $d_{z^2}$ order at a YBa2Cu3O7/ manganite interface

Roxana Gaina[1,2,*,†], Christopher W. Nicholson[1,*], Maxime Rumo[1], Subhrangsu Sarkar[1], Jarji Khmaladze[1], Eugenio Paris[3], Yi Tseng[3], Wenliang Zhang[3], Teguh C. Asmara[3], Daniel McNally[3], Cinthia Piamonteze[3], Eugen Weschke[4], Thorsten Schmitt[3], Claude Monney[1], and Christian Bernhard[1,†]

[1]University of Fribourg, Department of Physics and Fribourg Center for Nanomaterials, Chemin du Musée 3, CH-1700 Fribourg, Switzerland

[2]Laboratory for Neutron Scattering and Imaging, Paul Scherrer Institut, CH-5232 Villigen PSI, Switzerland

[3]Swiss Light Source, Photon Science Division, Paul Scherrer Institut, CH-5232 Villigen PSI, Switzerland

[4]Helmholtz-Zentrum Berlin für Materialen und Energie, BESSY II, D-12489 Berlin, Germany

[*]Equally contributing authors
[†]Correspondence to: christian.bernhard@unifr.ch; roxana.gaina@unifr.ch

**The interplay of nearly degenerate orders in quantum materials can lead to a myriad of emergent phases. A prominent case is that of the high-$T_c$ cuprates for which the relationship between superconductivity and a short-ranged, incommensurate charge density wave in the CuO2 planes involving the $d_{x^2-y^2}$ orbitals (Cu-CDW) is a subject of great current interest. Strong modifications of the strength and coherence of this Cu-CDW have been achieved by applying large magnetic fields, uniaxial pressure, or via the interfacial coupling in cuprate/manganite multilayers. However, such modifications do not alter the dominant orbital character. Here we investigate cuprate/manganite multilayers with resonant inelastic X-ray scattering (RIXS) and show that a new kind of Cu-based density wave order can be induced that has not been previously observed in the cuprates. This order has an unusually small wave vector of Q ≈ 0.096 reciprocal lattice units (r.l.u.), a large correlation length of about 40 nm, and a predominant $d_{z^2}$ orbital character, instead of the typical $d_{x^2-y^2}$ character. Its appearance is determined by the hole doping of the manganite which is a key parameter controlling the interfacial charge transfer and orbital reconstruction. We anticipate that the observation of a previously unknown type of density wave order at the YBCO interface will allow for fresh perspectives on the enigmatic relation between superconductivity and charge order (CO) in the cuprates.**

A. Introduction

The complex phase diagram of the cuprate high-$T_c$ superconductors is a prime example of the varied properties of strongly correlated quantum materials that arise from competing orders and nearly degenerate ground states.[1] Besides the enduring puzzle of the superconducting pairing mechanism, the exact nature of the normal state in these materials remains elusive. In particular, in the underdoped regime where a so-called pseudogap depletes the low energy electron states already well above $T_c$[2,3,4], various spin and/or charge orders have been found to coexist with superconductivity. Intense studies have been carried out on the remnants of antiferromagnetic (AF) correlations in the strongly underdoped regime[5,6] and on the so-called stripe order with coupled spin and charge modulations in $La_{1.875}Ba_{0.125}CuO_4$ [7,8]. More recently, the focus has shifted to an incommensurate Cu-CDW order that was first observed in underdoped $YBa_2Cu_3O_{7-\delta}$ (YBCO) [9,10], but meanwhile also in several other compounds. In bulk samples at ambient conditions this static Cu-CDW, with a wave vector of $Q_\parallel \approx 0.25-0.3$ r.l.u. and $d_{x^2-y^2}$ orbital character, is relatively weak and short-ranged. Its relationship with superconductivity remains controversial, with the proposed scenarios ranging from simple competition[11], over intertwined orders such as the enigmatic pair-density wave state[12,13], to a superconducting pairing mechanism that is mediated by CDW-fluctuations[14,15,16]. It is still debated whether the short-range CDW correlations are driven by Fermi-surface nesting[17], or rather by a tendency towards charge segregation provoked by strong electronic correlations[16]. However, the recent observation of a re-entrant Cu-CDW with a small wave vector of $Q_\parallel = 0.15$ r.l.u. in strongly overdoped Bi-2201 is difficult to reconcile with a pure nesting scenario[16].

Various successful attempts have been undertaken to enhance and modify the Cu-CDW in order to further explore its origin and relationship with superconductivity. In underdoped YBCO, the strength and the correlation length of the Cu-CDW with $Q_\parallel \approx 0.3$ r.l.u. have been strongly enhanced by applying large magnetic fields that suppress superconductivity[18,19] or uniaxial pressure along the a-axis, which enhances the orthorhombicity, but suppresses SC only partially[20]. In both cases, there appears to be a strong competition between SC and the long-range CDW. A strong enhancement of the Cu-CDW with $Q_\parallel \approx 0.3$ r.l.u. has also been achieved in YBCO/manganite multilayers for which the competition with superconductivity appears to be weak[21]. Moreover, for YBCO/ $Nd_{1-x}(Ca_{1-y}Sr_y)_xMnO_3$ (YBCO/NCSMO) multilayers with x=0.35, it has been demonstrated that the magnitude of the Cu-CDW can be strongly modified via the tolerance factor (or the Sr content) NCSMO, which controls its intrinsic charge and orbital order[22].

The YBCO layers of the latter YBCO/NCSMO multilayers have anomalous superconducting properties. For multilayers for which the tolerance factor, $t$, (or the Sr content, y) of the manganite is small enough to induce an insulating state with CE-type charge and orbital (Mn-CO) order, the YBCO layers exhibit a granular superconducting state which gives rise to an insulator-like upturn of the resistance below $T_c$, instead of the expected drop to zero resistance[22,23,24]. This unusual behavior occurs only in multilayers for which the thickness of the YBCO layer is below a critical value of $d^{YBCO} \leq 7-8$ nm and where the manganite layers display the Mn-CO[25,23]. Notably, a coherent SC response below $T_c$ can be restored here by application of a large magnetic field[25], which is known to weaken and suppress the Mn-CO of the manganite. This unusual kind of magnetic-field-induced transition from an insulating to a superconducting state is absent in corresponding NYN multilayers for which the manganite has a larger tolerance factor (Sr content, y) and thus an itinerant ferromagnetic instead of the Mn-CO ground state[24]. These trends suggest that the granular superconducting state of the YBCO layers is induced by a proximity effect due to the coupling with the Mn-CO of the adjacent manganite layers. While the interfacial coupling mechanism remains to be established, is seems likely that the concomitant induced (or strongly enhanced) Cu-CDW order in the YBCO layer plays an important role in the formation of the granular superconducting state.

This calls for further studies of the relationship between the manganite and cuprate charge/orbital orders in these multilayers. In the following, we present a combined resonant inelastic and elastic x-ray spectroscopy (RIXS and REXS) and x-ray absorption spectroscopy (XAS) study at the Cu $L_3$-edge of a $Nd_{1-x}(Ca_{1-y}Sr_y)_xMnO_3$/ $YBa_2Cu_3O_7$/ $Nd_{1-x}(Ca_{1-y}Sr_y)_xMnO_3$ (NYN) trilayer with x=0.5 and y=0.25. In this sample, the hole doping of the manganite has been adjusted to obtain a long-ranged Mn-CO order, unlike the previously studied multilayers with x=0.3-0.35 for which the Mn-CO is strongly disordered[24,25]. To our surprise we find that an entirely new kind of Cu-based order is induced in the YBCO layer below 170 K that has an unusually short wave vector of $Q_\parallel \approx 0.096$ r.l.u., a large correlation length of about 40 nm, and a predominant $d_{z^2}$ character, rather than the standard $d_{x^2-y^2}$ one. Moreover, we identify the hole doping of the manganite layers as a key parameter that determines not only its own charge/orbital and magnetic order but also allows modification of the interfacial charge transfer and the occupation of the $d_{z^2}$ orbitals in the interfacial $CuO_2$ planes. Our observations provide insights into the interfacial coupling between the Mn- and Cu-based charge and magnetic orders, and they suggest a new route to induce coupled quantum states in these cuprate/manganite multilayers.

### B. Experimental

**Different Bragg peaks of elastic RIXS signal at x=0.5 and 0.35**

Figure 1(a) shows a sketch of a NYN trilayer and the scattering geometry of the RIXS setup. The angle between the incident ($\mathbf{k_{in}}$) and scattered ($\mathbf{k_{out}}$) x-ray beams ($\Psi = 50°$), and the resulting magnitude of the scattering vector $\mathbf{Q}=\mathbf{k_{in}}-\mathbf{k_{out}}$ are kept fixed during the experiment. The in-plane component $Q_\parallel$ is varied here by rotating the sample, and is zero at specular reflection, when the angles α and β are equal (see Fig. 1a). The corresponding variation of the out-of-plane component $Q_\perp$ is not discussed here since the quasi-2D Cu-based order is expected to result in a broad peak in the direction perpendicular to the $CuO_2$ planes. Typical RIXS intensity spectra of the outgoing scattered photons as a function of their energy loss (ΔE) are displayed in Figure 1(b) for different in-plane momentum transfer values. In the present case, the incident energy is resonant with the Cu *L*-edge (i.e. resonant with a Cu *2p* to *3d* transition).

As a consequence, the elastic signal at ΔE=0 is very sensitive to the scattering from the Cu-*3d* valence electrons and is used in the following to probe the Bragg peak of the charge/spin order of the Cu ions. The inelastic signal at ΔE<0 is not further discussed here, but contains rich information on phonons, magnons and crystal field excitations between the Cu-*3d* levels that will be presented elsewhere. As can be clearly observed in the raw data, a significant enhancement of the elastic peak is found within a particular range of $Q_\parallel$ values, signifying the occurrence of a Bragg peak, and hence an underlying order.

The $Q_\parallel$ dependence of the elastic signal of the NYN trilayer with x=0.5 is analyzed in more detail in Figure 1(c), revealing a pronounced Bragg peak at an in-plane wave vector of $Q_\parallel \approx$ −0.096 r.l.u. To obtain this curve, the RIXS spectra in Fig. 1b at different in-plane scattering vectors are obtained by rotating the sample surface with respect to the beam. Each spectra is normalized to the integrated intensity of the *dd* excitations, and the elastic part of the RIXS spectra is then summed to give the total elastically scattered intensity for each wavevector. Interestingly, the observed wave vector is much smaller than the one of the Bragg peak due to the ordinary Cu-CDW of YBCO, which is expected at $Q_\parallel \approx \pm 0.3$ r.l.u [10]. The scan over a wider $Q_\parallel$-range in Figure 1(d) reveals that no Bragg peak around $Q_\parallel \approx 0.3$ r.l.u. occurs for the x=0.5 NYN trilayer. Also shown, for comparison, are the data of a NYN trilayer with x=0.35 and y=0.2 (green symbols) for which a pronounced Bragg peak is indeed observed at $Q_\parallel \approx -0.33$ r.l.u., as reported by Perret et al.[22]. Correspondingly, from Resonant Elastic X-ray Scattering (REXS) data (Supplementary Figure 4, SOM), the x=0.35 NYN trilayer shows no sign of a Bragg peak around $Q_\parallel \approx -0.096$ r.l.u. This observation of mutually exclusive Bragg peaks in the x=0.5 and x=0.35 NYN trilayers implies that the underlying incommensurate orders are of different nature and are controlled by the hole doping of the manganite layers.

Comparably small ordering wave vectors as observed in the x=0.5 NYN trilayer have only been reported for strongly overdoped Bi-2201 with $Q_\parallel \approx 0.15$ r.l.u.[16], and for the spin component of the so-called stripe order of LBCO with $Q_\parallel \approx 0.125$ r.l.u.[7,8]. The Bragg peak in the present case is also rather narrow with a value of $\Delta Q_\parallel \approx 0.003$ r.l.u. corresponding to a correlation length of $\xi = a\left(\pi \Delta Q_\parallel\right)^{-1} \approx 39$ nm (assuming the in-plane lattice parameter a=3.87 Å as given by the LSAT substrate). This is about an order of magnitude larger than the typical values of YBCO bulk or thin film samples, with the exception of the 3D-CDW that can be induced with high magnetic field[18,19] and uniaxial pressure[20].

### Polarization dependence of Bragg peak at x=0.5

Further evidence of the distinct nature of the Cu-based order underlying the Bragg peak at $Q_\parallel \approx 0.096$ r.l.u. comes from measurements obtained at different polarizations, and from comparing the enhancement of the peak obtained on either side of Γ-point of the Brillouin zone (Fig. 1d, inset). The elastic scattering intensity measured in a RIXS experiment depends on geometric factors and the orbital character of the states involved in the scattering process. Particularly relevant are the electric field component of the X-ray beams with respect to the scattering plane (π or σ polarization), and the incident angle of the beam with respect to the sample surface (grazing incidence or grazing exit beam, see Figure 1(a)). The latter of these defines the surface projection of the momentum transferred during the RIXS process, and hence the position in the Brillouin zone. The effect of these factors will clearly vary depending on the orbital symmetry involved in the scattering process e.g. in-plane $d_{x^2-y^2}$ or out-of-plane $d_{z^2}$. Therefore, obtaining measurements under different geometries and comparing with scattering cross-sections calculations for specific orbitals, allows the determination of the dominant orbital symmetry for a particular scattering process. Figures 2(a) and 2(b) show the $Q_\parallel$ scans at 930.4 eV for π and σ polarization in grazing incidence (incident angle $> 0$) and grazing exit (incident angle $< 0$) geometries. The overlaid fits of the Bragg peak are obtained using a Lorentzian line-shape and an exponential background; dashed grey lines show the background contribution. We observe that the Bragg peak enhancement is considerably larger for π-polarization than for σ-polarization for both grazing incidence and exit geometries. This corresponds to a complete reversal compared with previous reports of $d_{x^2-y^2}$ order[10,16], where σ-polarization shows the strongest signal. This strongly suggests that the behavior underlying the newly observed Bragg peak corresponds to a different orbital symmetry than in previous observations. This finding is further highlighted in Figure 2(c), which compares the maximum Bragg peak intensities (symbols) obtained under the different scattering conditions presented in Fig. 2(a) and (b) with the corresponding calculated RIXS cross-sections as a function of δ. By comparing the expectations of the RIXS cross-sections for a $d_{x^2-y^2}$ orbital under the same scattering conditions as in the experiment (see Methods and Ref.[26]), it is immediately clear that the observed geometric dependence cannot be explained by a dominant $d_{x^2-y^2}$ orbital character. This highlights that the order with Q = 0.096 r.l.u. does not have the $d_{x^2-y^2}$ orbital character typically observed in the cuprates.

Instead, the enhancement of the Bragg peak intensity with out-of-plane light polarization indicates the participation of $d_{z^2}$ orbitals. Indeed, we find a reasonable agreement with the calculated RIXS cross-sections for scattering into $d_{z^2}$ states with a spin-flip process. The calculated curves for spin-flip processes with spins oriented along [001] are shown in Figure 2 (d), and are again compared with the Bragg peak intensities under different scattering conditions. In particular, the spin-flip scattering involving $d_{z^2}$ orbitals correctly predicts the observed reversal of cross sections compared with the $d_{x^2-y^2}$ case in Fig. 2 (c). This therefore implies a dominant role of the out-of-plane $d_{z^2}$ orbitals in the Q = 0.096 r.l.u. order. In the supplementary materials, we present comparisons between additional orbital symmetries and the observed geometric dependence. These reveal that $d_{z^2}$ orbitals with spins pointing along either [001] or [110] can reproduce the reversed cross-section behaviour with respect to $d_{x^2-y^2}$ orbitals. Furthermore, the calculations highlight that no configuration with in-plane $d_{x^2-y^2}$ orbitals comes close to reproducing the observed polarization and angular-dependence of the Bragg peaks. This gives additional confidence in the assignment of $d_{z^2}$ character in the new order. Differentiating the possible spin scenarios, and whether there is mixed charge and spin character to the order, will require additional experimental input, such as a polarization analysis of the scattered X-rays that goes beyond the scope of the present work.

The present results therefore establish that the x=0.5 NYN trilayer hosts a new kind of incommensurate Cu-based order that strongly involves the $d_{z^2}$ orbitals and likely has a mixed spin/charge character. In the following, we denote it for simplicity as Cu-$d_{z^2}$ order.

**Resonant Elastic X-ray Scattering (REXS) study of the Bragg peak at x=0.5**

The Bragg peak of the x=0.5 NYN trilayer due to the Cu density wave order at $Q_{\parallel}$ ≈ −0.096 r.l.u. and its unusual polarization dependence have also been observed with Resonant Elastic X-rays Scattering (REXS) experiments[27,28] for which the incident photon energy has been set to the Cu-$L_3$ edge and tuned to the maximum of the intensity of the XAS signal, i.e. to 931.2 eV. In a REXS experiment the energy of the diffracted x-rays is not analyzed, the Bragg peak of the elastic signal is therefore superimposed on a background that includes the entire inelastic signal and thus is larger than in RIXS. The scattering geometry of the REXS experiment was adjusted to probe the Bragg peak in the same configuration as in the RIXS experiments, i.e. at (H K L) = (-0.096 0 1.56). The value of H was scanned while keeping K and L constant by rotating the sample as well as the detector to vary the angle of incidence on the sample θ and the scattering angle 2θ. We have also confirmed that the intensity of the Bragg peak at H ≈ −0.096 r.l.u. is only weakly dependent on L.

Figure 3 displays a wide-range H-scan at 9 K for both π and σ polarizations, which confirms that the Bragg peak due to the Cu-$d_{z^2}$ order occurs on both sides of the specular peak at (±0.096 0 1.56) and has a much higher intensity for π-polarization. Moreover, it shows for both polarizations how the background has been accounted for with a polynomial function. Figure 3(b) displays the background subtracted response for both polarizations, which highlights a large difference between π- and σ-polarization in the intensity of the Bragg peaks at H ≈ ±0.096 r.l.u that confirms the trend seen in the RIXS data evidencing a predominant $d_{z^2}$ orbital character of this Cu density wave order, as discussed in the previous section and shown in Figure 2.

Figures 3(c)−3(e) display representative curves of a temperature scan that was performed in π-polarization around the (-0.096 0 1.56) Bragg peak. The background has been described with a simpler, linear function that also provides a reasonable description of the background of the full scan around the specular peak (π-polarization, in Figure 3(a)). Both the Bragg peak due to the charge order and the specular peak at H=0 have been described with Gaussian functions. The temperature dependence of the obtained position and area of the Bragg peak is displayed in Figures 3(f)−3(g). It reveals that the Bragg peak due to the static Cu-$d_{z^2}$ order persists to rather high temperature and vanishes only above 170 K similar to the usual $Q \approx 0.3$ r.l.u order[10]. The Cu-$d_{z^2}$ order thus has a somewhat higher onset temperature than the Cu-CDW with $d_{x^2-y^2}$ character, which develops rather gradually below about 150 K in bulk YBCO[10,11] and in the range from 120 to 175 K in corresponding YBCO/manganite heterostructures with x=0.3−0.35[21,22].

### XAS and resonance of RIXS Bragg peak

We have complemented the above results with X-ray absorption spectroscopy (XAS) measurements that reveal important differences between the NYN trilayers for x=0.5 and x=0.35 with respect to the strength of their interfacial charge transfer and orbital reconstruction.

Figure 4(a) shows a sketch of the XAS experiment for which X-rays with π and σ polarization at an incidence angle of 30° are absorbed by the sample. Information on the interfacial Cu ions and the more bulk-like Cu ions is obtained from the comparison of the XAS spectra measured in total electron yield (TEY) and fluorescence yield (FY), respectively. Due to the very small escape depth of the photo-excited electrons of only a few nanometers, the TEY signal arises here predominantly from the Cu ions at the topmost YBCO/manganite interface. The FY signal, however, is equally sensitive to all Cu ions and thus representative of the majority of the Cu ions away from the interface.

Figures 4(b) and (e) compare the FY spectra of the trilayers with x=0.35 and x=0.5. These resemble the spectra of bulk-like YBCO showing a sharp resonance at 931 eV and a strong linear dichroism towards the σ polarization ($\mu_{ab}$) that confirms that the holes in the CuO$_2$ planes away from the interface have predominantly $d_{x^2-y^2}$ character. The corresponding TEY signal of the x=0.35 trilayer in Figure 4(c) reveals some remarkable differences with respect to the FY signals. In particular, a clear redshift of the resonance by about 0.5 eV is observed, as well as a strong enhancement of the c-axis component $\mu_c$ with π−polarisation that gives rise to a weak inversion of the X-ray linear dichroism (XLD). Both features were previously observed in YBCO/manganite multilayers with a similar hole doping of the manganite of x=0.3−0.35[29]. Such changes have been explained in terms of a transfer of electrons from the manganite to the YBCO and an orbital reconstruction of the interfacial Cu ions that leads to a redistribution of holes from the $d_{x^2-y^2}$ to the $d_{z^2}$ orbitals[30,31]. Moreover, it has been shown that this charge transfer and orbital reconstruction are robust features that occur at this doping level irrespective of whether the manganite layer is an itinerant ferromagnet[30,31] or an antiferromagnetic charge/orbital ordered insulator[22].

The more remarkable is our finding that for the TEY resonance of the trilayer with x=0.5 in Figure 4(d) both the red shift and the enhancement of the c-axis signal, and thus the interfacial charge transfer and orbital reconstruction are strongly reduced compared to the x=0.35 sample in Figure 4(c). The reduced transfer of electrons from the manganite to the YBCO layer (or the smaller red shift of the TEY resonance) is naturally explained in terms of the larger hole doping of the manganite at x=0.5, which lowers the Fermi-level of the manganite and thereby reduces its mismatch with respect to that of YBCO. The reason for the concomitant weakening of the orbital reconstruction of the interfacial Cu ions (i.e. a reduced density of $d_{z^2}$ holes) is less obvious. It may involve additional effects such as a change of the position of the apical oxygen ion, which slightly reduces the hybridization of the $d_{z^2}$ levels of the interfacial Mn and Cu ions. A full understanding requires further theoretical and experimental studies of the detailed structure of the interface, which hopefully will be motivated by our work.

Finally, Figure 4(f) shows the evolution of the amplitude of the Bragg peak at $Q_\parallel \approx -0.096$ r.l.u. in the elastic RIXS data at x=0.5 as a function of the incident photon energy (normalized to the intensity of the *dd*-excitations). It exhibits a pronounced resonance with a maximum around 930.4 eV that is close to the resonance of the XAS signal in TEY mode (Figure 4(d)) and clearly smaller than the value of the FY mode at 931eV (Figure 4(e)). This coincidence with the resonance of the TEY signal agrees with our interpretation that this new kind of Cu-$d_{z^2}$ order originates predominantly from the interfacial $CuO_2$ planes. Certainly, it shows that the $Q_\parallel \approx -0.096$ r.l.u. order originates from the $CuO_2$ planes and excludes the possibility that it involves the Cu ions of the CuO chains, for which the resonance would occur at a much higher energy of about 933.8 eV[32,33]. Moreover, a Bragg-peak that arises from a superstructure of oxygen vacancies in the CuO chains would persist to much higher temperature (well above room temperature) in clear contrast to the temperature dependence shown in Fig. 3(g).

### C. Discussion

The present work emphasizes the range of ordered phases that develop in the high-$T_c$ cuprates, and, in particular, suggests that these can be induced to involve the out-of-plane orbitals, in contrast to previous observations. This raises important questions about the coupling mechanism at the cuprate/manganite interface that induces (or enhances) these different orders. It is likely that Jahn-Teller distortions due to the CE-type charge/orbital order of the manganite are induced that can be transported into the cuprate via the interfacial Ba-O layer. In turn, these distortions may trigger an orbital order of the interfacial $CuO_2$ layer through a variation of the orbital reconstruction along the interface with a certain alternation of the $d_{z^2}$ and $d_{x^2-y^2}$ levels. A lateral ordering of the $d_{z^2}$ and $d_{x^2-y^2}$ orbitals could also explain the magnetic component of the Cu-$d_{z^2}$ order that is suggested by our calculations. It would give rise to a modulation of the in-plane antiferromagnetic exchange couplings since the exchange between a pair of $d_{x^2-y^2}$ spins is strongly reduced when one orbital is replaced by a $d_{z^2}$ orbital, and even becomes weakly ferromagnetic for spins on two adjacent $d_{z^2}$ orbitals[31]. Additionally, the Cu-$d_{z^2}$ order in the multilayer with x=0.5 occurs despite a reduced density of $d_{z^2}$ holes on the interfacial $CuO_2$ plane as compared to the multilayer with x=0.35, for which the usual $d_{x^2-y^2}$ CDW order prevails. This suggests that an optimal density of $d_{z^2}$ holes, or a particular ratio with respect to the $d_{x^2-y^2}$ holes, is required to stabilize the Cu-$d_{z^2}$ order and might be directly linked to its particularly long wavelength (small wave vector).

Another important aspect concerns the relationship of this new Cu-based density wave order with superconductivity. The magneto-transport data of the NYN trilayer with x=0.5, that are detailed in the SOM and reported in Khmaladze et al.[24], provide evidence for a strong granularity of the superconducting order. They also suggest that the grain boundaries, which break the superconducting coherence and lead to Coulomb-blocking effects in the YBCO layer, are related to the domain boundaries of the Mn-CO in the manganite layers. Indeed, the Coulomb-blockade is stronger in NYN trilayers with x=0.35, for which the Mn-CO order is weaker and more short-ranged (high density of domain-boundaries), than in corresponding NYN trilayers with x=0.5 (with fewer domain boundaries)[24]. Interestingly, the induced Cu-based density wave order follows a similar trend, i.e. the correlation length of the Q ≈ −0.096 r.l.u. order at x=0.5 with ξ ≈ 40 nm is significantly larger than the one of the Cu-CDW with Q~0.3 r.l.u. at x=0.35 with ξ < 10 nm. This fits into the picture that the Cu-based density wave order acts as a mediator between the Mn-CO and superconductivity and suggests that the phase coherence of the SC order may be linked to the one of the Cu-based density wave order.

Our findings open new routes for exploring the physics of intertwined ordered phases and demands both further theoretical and experimental work on the electronic and structural properties of the cuprate/manganite interface. They also raise hopes that other exotic Cu-based orders can be induced by modifying the doping of the cuprate and manganite layers. In particular, the manganite provides a range of charge, orbital and magnetic orders for which the interaction with the cuprate remains unexplored. Finally, with electric field gating it may even be possible to switch between different Cu-based orders at the YBCO/manganite interface to study their relationship with superconductivity or to develop new device and sensing applications.

**Methods**

**Sample growth and characterization**.

Trilayers of $Nd_{1-x}(Ca_{1-y}Sr_y)_xMnO_3$(20nm)/ $YBa_2Cu_3O_7$(7nm)/ $Nd_{1-x}(Ca_{1-y}Sr_y)_xMnO_3$(10nm) with x=0.5, y=0.25 and x=0.35, y=0.2 and a 1 nm thick $LaAlO_3$ capping layer were grown with pulsed laser deposition (PLD) on (001)-oriented $La_{0.3}Sr_{0.7}Al_{0.65}Ta_{0.35}O_3$ (LSAT) substrates. Details of the growth conditions and the sample characterization are given in the Supplementary Notes in section 1. The electronic transport and magnetization measurements showing the superconducting and magnetic properties of the trilayers are reported in Ref.[24]. In particular, for the NYN trilayer with x=0.5 and y=0.25 the R-T curve in zero magnetic field exhibits the onset of superconducting transition below $T_c$≈82 K followed by a steep increase of the resistance below about 60 K that is characteristic of a granular SC state. In 9 Tesla a regular superconducting transition occurs with the same onset temperature that is completed below about 55K. Note that the SC transition, which for fully oxygenated and thus slightly overdoped bulk YBCO is around 90 K, is typically reduced for very thin YBCO layers. In bare thin films this is due to finite size and strain effects as well as related defects. For the present multilayers, there is in addition a reduction of the hole doping of the interfacial $CuO_2$ layers that arises due to the transfer of electrons from the neighboring manganite layers and the missing CuO chain layer at both the top and bottom interfaces[34,31,35,36].

**Resonant inelastic x-ray scattering.**

The RIXS experiments at the Cu $L_3$-edge were performed at the ADRESS beamline of the Swiss Light Source (SLS) at the Paul Scherrer Institute (PSI) in Switzerland[37,38]. The scattering angle between the incoming and outgoing X-ray beams was 130°. An energy resolution of 130 meV (full width at half-maximum) was determined from the elastic scattering off a carbon-filled acrylic tape. The incident photon energy scale was shifted in order to match the X-ray absorption spectrum measured in total electron yield (TEY) at ADRESS beamline to the corresponding spectrum obtained at the XTreme beamline (see below). The energy scale of the incident energies used for the RIXS data has been corrected accordingly.

**RIXS cross-section calculations.**

Cross-section calculations for the RIXS process were performed using a single-ion model following Moretti Sala *et al*[29]. This calculation takes into account the atomic symmetries of the core-level *p* and valence *d*-orbitals and the geometry of the experimental setup, namely the π and σ polarization of incident light and the scattering angles of the experiment. Since measurements were performed at the Cu *L*-edge, the atomic cross sections of the elastic RIXS signal were calculated assuming X-ray absorption and emission to and from valence *d*-levels. We show here calculations for valence states with symmetry $d_{x^2-y^2}$ or $d_{z^2}$. Calculations for other *d*-orbitals have been performed, but are found not to correspond to the obtained data. (Supplementary Note 4). Processes without a spin-flip (elastic) and with spin-flip have both been calculated, assuming spin along the [110] direction for $d_{x^2-y^2}$ states and along both [001] and [110] for $d_{z^2}$ states. In order to compare the calculated cross-section with the observed RIXS intensities (arbitrary units), a constant conversion factor is applied, which is given in each of the figure captions.

**Resonant Elastic X-ray Scattering (REXS)**

The REXS experiments at the Cu $L_3$-edge were performed at the UE46 PGM-1 beamline of the BESSY II synchrotron at Helmholtz-Zentrum Berlin fur Materialen und Energie (HZB) in Germany. The X-ray photodiode from the diffractometer can move in a continuous way to a large range of angles and the energy-integrating nature of the photodiode will show the elastic peak superimposed on a background that includes all the inelastic components. The scattering geometry of the REXS was adapted to probe the Bragg peak in similar configuration as RIXS experiments.

**X-ray absorption spectroscopy**

X-ray absorption spectroscopy (XAS) and x-ray linear dichroism (XLD) at the at the Cu $L_3$-edge were measured in total electron yield (TEY) and fluorescence yield (FY) mode at the XTreme beamline of the SLS at the PSI in Switzerland.[39] The angle of incidence of the X-ray beam was set to 30° such that the response in σ polarization, $\mu_\sigma$, corresponds to the ab plane response of YBCO, $\mu_{ab} = \mu_\sigma$, whereas the c-axis response, $\mu_c$, is obtained from the signal in π-polarization, $\mu_\pi = \mu_c{}^*$ according to the geometrical factor, $\mu_c = 1/(\mu_\pi \cos^2(\theta) - \mu_\sigma \tan^2(\theta))$.

In Figure 3 (b) - (e) the data have been normalized to the maximum of the polarisation-averaged absorption $\mu_{avg} = \max[(2\mu_{ab})/3]$ at the Cu $L_3$-edge.

**Acknowledgments**
We acknowledge fruitful discussion with Dr. Steven Johnston and experimental support provided by Dr. Enrico Schierle during the REXS beamtime. Work at the University of Fribourg was supported by the Swiss National Science Foundation (SNSF) through Grants No. 200020-172611 and CRSII2-154410/1. C.M. acknowledges the support by the SNSF grants No. PZ00P2_154867 and PP00P2_170597. The RIXS and XAS experiments were performed at the ADRESS beamline respectively the Xtreme beamline of the Swiss Light Source at the Paul Scherrer Institute. The work at the ADRESS beamline is supported by the Swiss National Science Foundation through project no. 200021_178867, the NCCR MARVEL and the Sinergia network Mott Physics Beyond the Heisenberg Model (MPBH). The REXS experiments (Supplementary information) were performed at UE46_PGM-1 beamline at the BESSY II synchrotron in the Helmholtz-Zentrum, Berlin (HZB). This project received funding from the European Union's Horizon 2020 research and innovation program under grant agreement No 730872. We thank J-L. Andrey and O. Raetzo from the Physics Department workshop at UniFr for technical assistance.

**Author contributions**
RIXS measurements were carried out by R.G., C.W.N., M.R, S.S., E.P., Y.T., W.Z., T.C.A., T.S., C.M. and C.B. and analyzed by R.G. and C.W.N. Samples were grown and characterized by J.K. and S.S.. Experimental support during the RIXS experiment, including alignment of the spectrometer and maintenance of the beamline were performed by E.P., Y.T., W.Z., D.M, T.C.A. and T.S.. XAS measurements were obtained by R.G. and C.P. and analyzed by R.G.. The RIXS calculations were performed by C.W.N and C.M.. The REXS measurements were carried out by R.G., S.S., C.B. and E.W. and analyzed by R.G. and S.S.. The manuscript was written by C.W.N., C.M., R.G. and C.B. with contributions from all authors. C.M. and C.B. were responsible for the overall project planning and direction.

**Additional information**
Supplementary information is available in the online version of the paper.

**Competing financial interests.**
The authors declare no competing financial interests.

**Data availability**
The data will be freely available upon reasonable request to the corresponding authors.

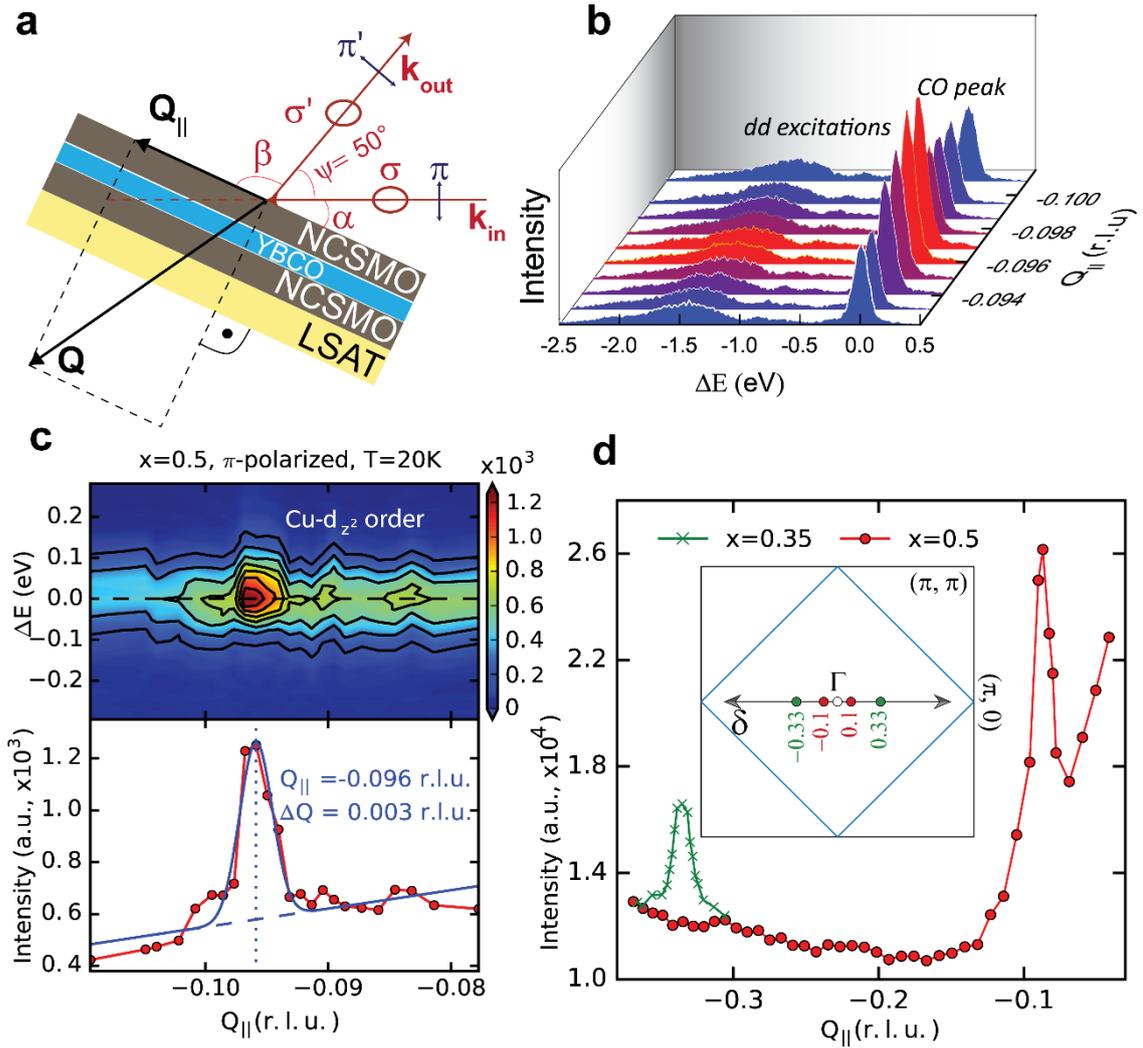

Figure **Error! Main Document Only.** : *Overview of the incommensurate Cu density wave orders in NYN trilayers obtained with RIXS.* **(a)** Schematic of the RIXS experiment at the Cu L-edge with X-rays with linear σ- and π-polarization. The scattering angle between in and outgoing beams is $\Psi = 50°$. The angle δ, which gives the projection of transferred momentum along the surface direction, is defined as zero when the angles α and β are equal (specular reflection) i.e. $\delta = 90° - \alpha - \Psi/2$. Therefore, grazing incidence corresponds to δ > 0, while grazing exit corresponds to δ < 0. **(b)** Observed RIXS spectrum for different $Q_{||}$ at the Cu $L_3$-edge for the trilayer indicating the elastic peak (ΔE = 0) enhancement due to CO and inelastic signals (ΔE < 0). **(c)** Evolution of the elastic peak intensity (ΔE = 0) in π-polarization as a function of the in-plane momentum transfer, $Q_{||}$, showing a pronounced Bragg peak at $Q_{||} \approx 0.096$ r.l.u. due to a Cu-based order. **(d)** Wide range $Q_{||}$-scan showing that the ordinary Cu-CDW Bragg peak around $Q_{||} \approx 0.33$ r.l.u. is absent for the trilayer with x=0.5. Shown for comparison are RIXS data of a trilayer with x=0.35 for which this latter Bragg peak is present and enhanced with respect to bulk YBCO (adapted from Ref [22]). The inset shows a schematic of the Brillouin zone and the positions of the two charge order peaks, which appear symmetrically about the Γ-point along the (π, 0) direction.

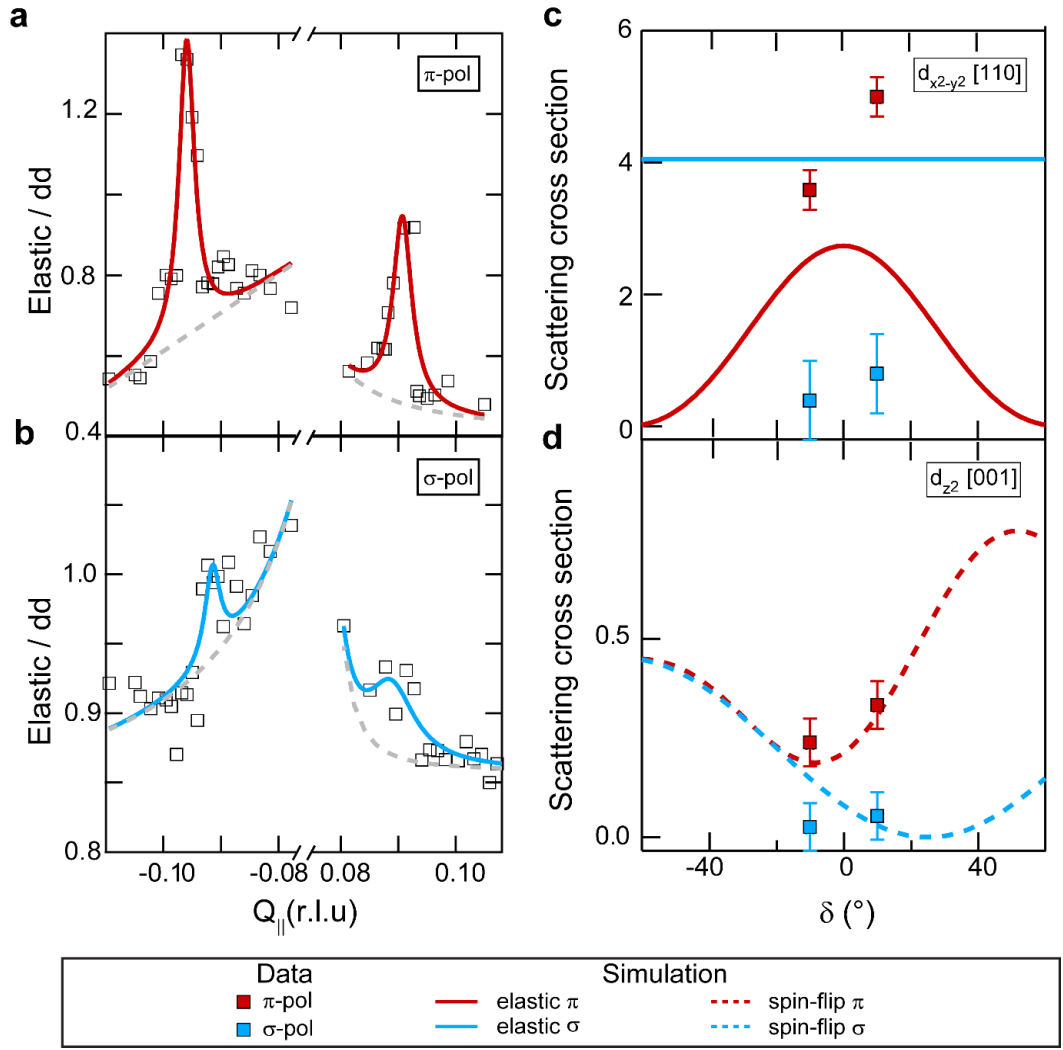

*Figure 2: **Evidence for the $d_{z^2}$ character of the Bragg peak of the x=0.5 NYN-trilayer obtained at the Cu $L_3$ edge. (a)** and **(b)** Normalized intensity of the elastic line from RIXS (open symbols) as a function of in-plane momentum transfer, $Q_{\parallel}$, of the incident X-rays for π- and σ-polarization respectively. Solid lines are Lorentzian fits with a Gaussian background. The grey dashed lines highlight the background contribution. The increased enhancement for the π-polarized beam are in direct contrast to previous observations of charge order involving $d_{x^2-y^2}$ orbitals. The combined uncertainty in $Q_{\parallel}$ is ±0.005 r.l.u.. **(c)** Geometric scattering cross-section calculations (solid lines) for an elastic $d_{x^2-y^2}$ process as a function of angle δ. An angle of 0° corresponds to the Γ-point. The intensities of the Bragg peaks at the corresponding scattering conditions (obtained for each of the four cases in (a) and (b)) are shown as markers (x30 magnified). The reversed polarization dependence compared with the calculations reveals the underlying orbital character cannot be $d_{x^2-y^2}$. **(d)** The same calculation as in (c) for a spin-flip scattering process involving $d_{z^2}$ orbitals, and spin oriented along the [001] direction. The $d_{z^2}$ orbital symmetry correctly captures the reversal of the measured π and σ-cross sections (markers, x2 magnified) compared with the $d_{x^2-y^2}$ case, highlighting the role of $d_{z^2}$ orbitals in the observed order.*

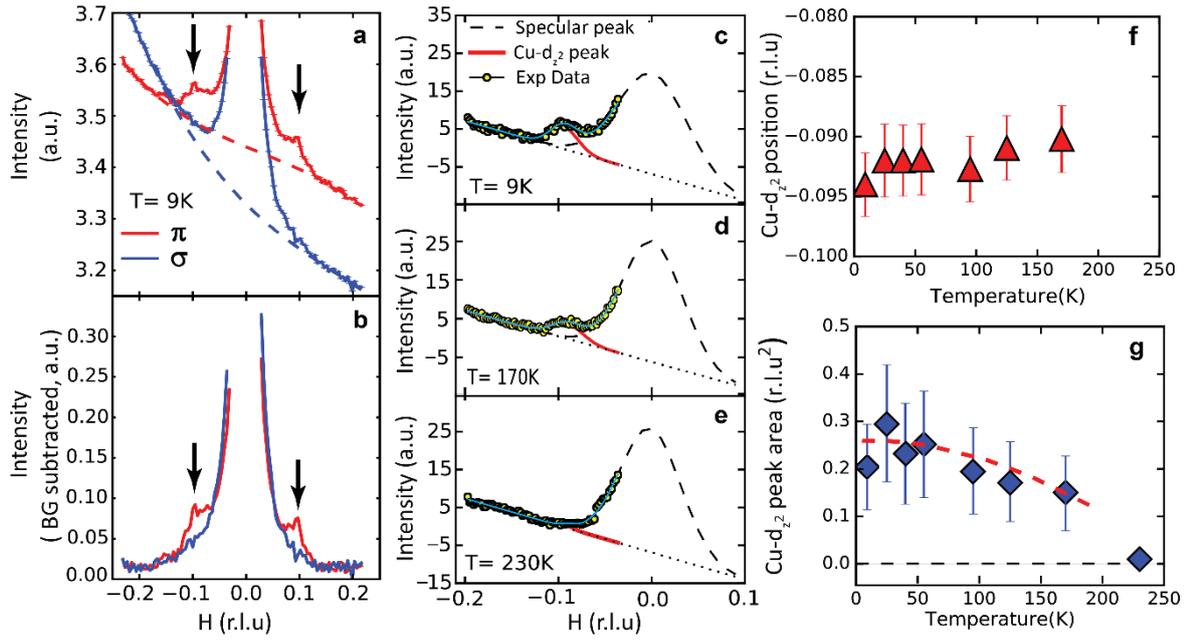

*Figure 3. **Summary of a REXS study of the charge order peaks in a NYN trilayer with x=0.5, y=0.25**. **(a)** Wide-range H-scan at 9 K which confirms that the Bragg peak due to the Cu-$d_{z^2}$ order is seen on both sides of the specular peak. **(b)** Background subtracted scans showing the strong polarization dependence of the Bragg peak due to the Cu-$d_{z^2}$ order that are marked by black arrows. **(c)-(e)** Partial scans of the Bragg peak around (-0.096 0 1.56) for selected temperatures of 9 K, 170 K, and 230 K, respectively. For each temperature, the plotted experimental data (open circles) represent the average of several consecutively acquired scans. The solid cyan lines denote the total fit. The background (described with a linear function) is shown as dotted line and the specular peak (Gaussian function) with a dashed line. The obtained Bragg peak of the Cu-$d_{z^2}$ order is displayed by a solid red line. Graphs **(f)** and **(g)** respectively, display the obtained temperature dependence of position and area of the Bragg peak of the Cu-$dz^2$ order. In **(g)**, the dashed red line is a guide to the eye. The propagation of uncertainty in peak position and area were calculated from the results of the fitting using standard procedures[39].*

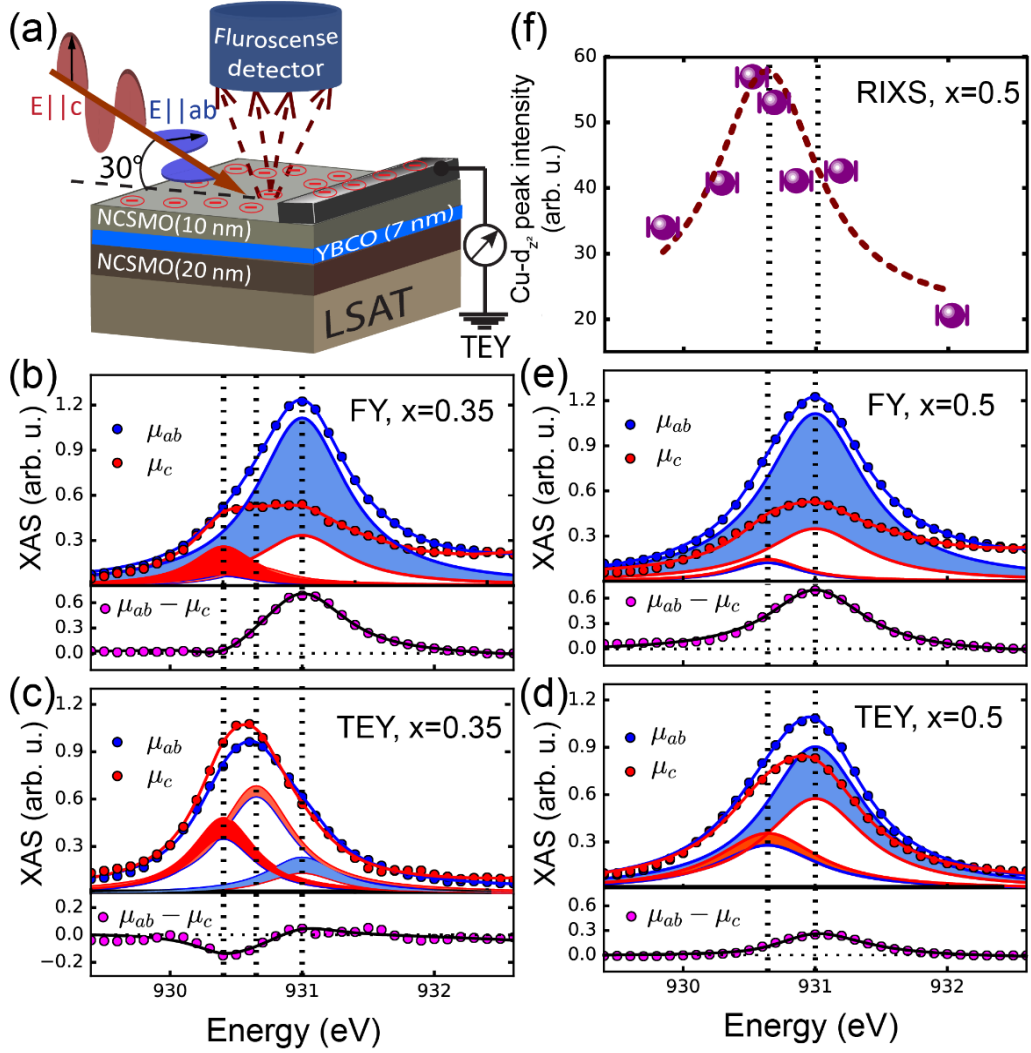

*Figure 4: **Charge transfer and orbital reconstruction probed with XAS.** (a) Schematic of the setup for measuring XAS in FY and TEY modes. (b) to (e) Comparison of the XAS spectra at the Cu $L_3$-edge at 20 K and their x-ray linear dichroism (XLD) for NYN trilayers with x=0.35 and x=0.5. The FY spectra in (b) and (e) represent the response of all Cu ions whereas the TEY spectra in (c) and (d) are governed by the interfacial Cu ions. Solid symbols show the experimental data and solid lines are fitted Lorentzian functions. The peak at 931 eV arises from the bulk-like Cu ions whereas the peaks at 930.3 and 930.6 eV are attributed to the interfacial Cu ions. The redshift of the latter with respect to the bulk-like resonance is due to the transfer of electrons across the LCMO/YBCO interface. The sign of the x-ray linear dichroism (XLD) of the peaks is shown by the colored shading: when $\mu_{ab} > \mu_c$ ( $\mu_{ab} < \mu_c$), the difference is shaded blue (red). The spectra reveal that the charge transfer and the orbital reconstruction are considerably stronger at x=0.35 than at x=0.5. (f) Evolution of the intensity of the Bragg peak in the elastic RIXS signal at $Q_{\parallel} \approx 0.096$ r.l.u. of the NYN trilayer with x=0.5 as a function of the incident photon energy around the Cu $L_3$-edge showing that the maximum coincides with the resonance of the interfacial Cu ions in the TEY-XAS signal (see panel **(e)**).*

# Supplementary Information

**Supplementary Note 1: Growth and characterization of manganite/cuprate heterostructures:**

*Sample Growth:* Trilayers for which a YBa$_2$Cu$_3$O$_7$ (YBCO) layer is embedded between two Nd$_x$(Ca$_{1-y}$Sr$_y$)$_x$MnO$_3$ (NCSMO) layers (NYN trilayers) along with a 2nm thick LaAlO$_3$ (LAO)

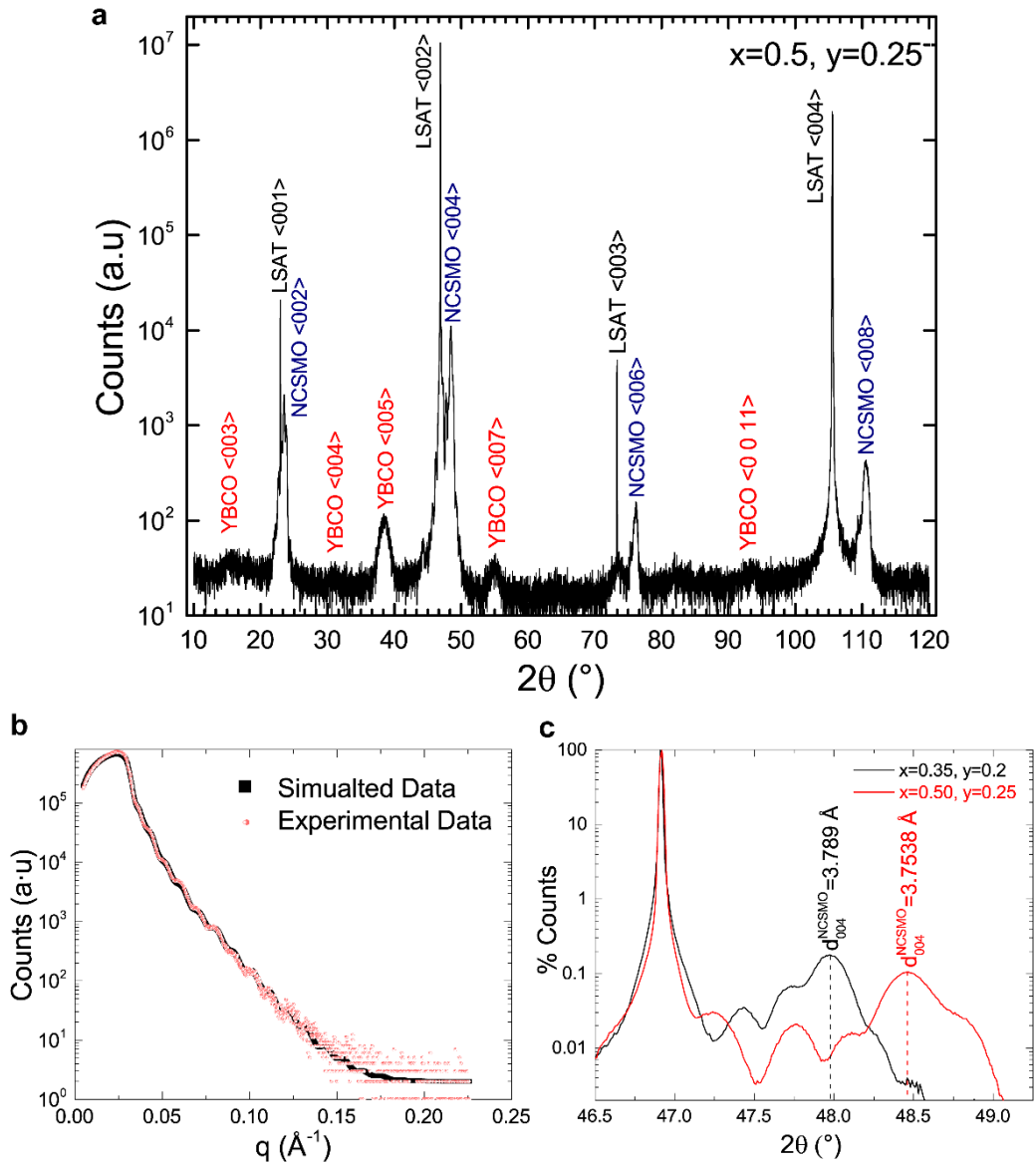

**Supplementary Figure 1**. *Structural characterization with x-ray diffraction and reflectometry. **(a)** $\theta - 2\theta$ scan of the NYN trilayer with x=0.5 with the assignment of the Bragg peaks to the YBCO and NCSMO layers and the LSAT substrate. **(b)** X-ray reflectivity curve of the NYN trilayer with x=0.5 (red symbols) and a fit (black line) from which the layer thicknesses have been obtained. **(c)** Magnified view of the $\theta - 2\theta$ scans around the (004) peak of NCSMO for the two NYN trilayers with x=0.5 and x=0.35 from which the out-of-plane lattice parameters have been deduced.*

capping layer were grown with pulsed laser deposition (PLD) on single crystalline (001)-oriented (LaAlO$_3$)$_{0.3}$(Sr$_2$TaAlO$_6$)$_{0.7}$ (LSAT) substrates (from Crystec). We used an excimer KrF laser ($\lambda$=248 nm, ts = 25 ns) with a spot size of the laser beam on the target of about 3

mm$^2$, a repetition rate of 2 Hz and a fluency of 1.42 J cm$^{-2}$. The LSAT substrates were placed about 5 cm above the targets and heated with an infrared laser from the backside at a rate of 20°C/min to the deposition temperature of 825°C. The substrates were preheated for at least one hour before the deposition was started. During the deposition the pressure of the oxygen gas with a purity of 99.5% was maintained at 0.35 mbar. After finishing the growth, the chamber was flooded with pure oxygen gas to generate near atmospheric pressure and the sample was cooled to 480°C at a rate of 10°C/min. Subsequently, the sample was annealed for about 1 hour between 480°C and 380°C to fully oxygenate the CuO chains of the YBCO layer before it was rapidly cooled to room temperature.

*Sample Characterization:* The structural properties of the NYN trilayers with x=0.5 and x=0.35 have been characterized with x-ray diffraction (XRD) and reflectivity (XRR) measurements using a Rigaku SmartLab (9kw) four-circle X-Ray diffractometer with a Cu-Kα$_1$ source and an incident parallel beam optics that consists of a reflecting mirror and a two-bounce Ge (220) monochromator ($\Delta\lambda/\lambda = 3.8 \times 10^{-4}$). Supplementary Figure **1(a)** shows the θ-2θ scan of the NYN trilayer with x=0.5 which confirms the epitaxial growth of the layers with the c-axis of the YBCO layer oriented along the surface normal. From the position of the assigned Bragg peaks the out-of-plane lattice parameters of 3.754 Å (pseudo-cubic) for NCSMO and 11.680 Å for YBCO are deduced. Supplementary Figure **1(b)** shows the X-ray reflectivity curve (red symbols) of the NYN trilayer with x=0.5 together with the best fit using the GenX software (black line) which yields layer thicknesses of 8.6 nm, 8.16 nm and 19.87 nm for the top NCSMO, middle YBCO and bottom NCSMO layers, respectively. Supplementary Figure **1(c)** shows a comparison of the θ-2θ scan around the (004) peak of NCSMO for the NYN trilayers with x=0.5 and x=0.35 which highlights that the former has a slightly smaller out-of-plane lattice parameter as is expected for the epitaxial growth on a LSAT substrate with a lattice parameter of 3.87 Å that gives rise to tensile strain conditions. Further details about the structural characterization of the NYN trilayer with x=0.35 can be found in Ref [1].

**Supplementary Note 2: R-T curves**

Supplementary Figure 2 compares the R-T curves of the NYN trilayer with x=0.5 and y=0.25 to the one of a single 7nm thick YBCO layer and a corresponding single 20nm thick NCSMO layer that were both grown under the same condition as the trilayer. The same data were already reported in Ref.[2]. Notably, the 7nm thick YBCO layer exhibits a regular superconducting transition at $T_C \approx 80$ K with no sign of an insulator-like upturn of the resistance below $T_C$ (blue line, panel (a)). The single manganite layer in panel (b) reveals a steep insulator-like upturn in zero field that is hardly altered even at 9T confirming the robustness of the Mn charge/orbital order. The R-T curve of the NYN trilayer at zero magnetic field (red line, panel (a)) exhibits the onset of a regular superconducting transition around 75 K that remains incomplete since it is sharply interrupted by a steep resistive upturn below about 60 K. At 9 Tesla (green line, panel (a)) the trilayer recovers a regular superconducting transition (that is however somewhat broader than the one of the 7nm thick YBCO layer).

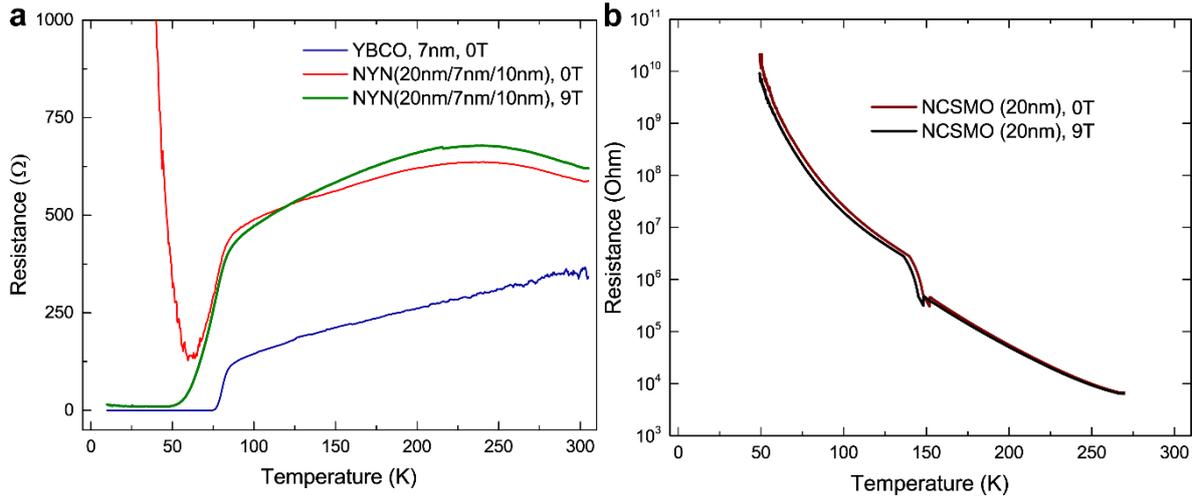

*Supplementary Figure 2: (a) Comparison of the R-T curves in 0 and 9 Tesla of the NYN trilayer with x=0.5 and y=0.25 with the one of a 7nm thick single YBCO layer at 0 T. (b) Corresponding R-T curves at 0 and 9 T for a single 20 nm thick NCSMO layer with x=0.5 and y=0.25.*

**Supplementary Note 3: Calculation of scattering cross sections**

Cross-section calculations for the RIXS process were performed using a single-ion model as described in the main text. This calculation takes into account the atomic symmetries of the core-level $p$ and valence $d$-orbitals and the geometry of the experimental setup, namely the π and σ polarization of incident light and the scattering angles of the experiment. Since measurements were performed at the Cu $L$-edge, the atomic cross sections of the elastic RIXS signal were calculated assuming X-ray absorption and emission to and from valence $d$-levels. We calculate the matrix element for elastic scattering as the product of two transitions: between the core-level state $2p$ to a valence state $3d$ and de-excitation from the same $3d$ state back to $2p$ state. Selection rules involving the addition of angular momenta are computed using the Wigner 3-$j$ symbols. In order to compare the calculated cross-section with the observed RIXS intensities (arbitrary units), a constant conversion factor is applied, which is given in each of the figure captions.

In the main text (Figure 2), it was shown that the angular dependence of the charge order enhancement could not be reproduced by assuming elastic scattering in the $d_{x^2-y^2}$ orbitals as the dominant scattering channel. In Fig. S3 a) we show that spin-flip scattering in $d_{x^2-y^2}$ orbitals similarly fails to capture the particular angular dependence observed in the experiments. In comparison, we find that the angular dependent intensity of the elastic background below the Bragg peak (the grey dashed lines in Figure 2) agrees very well with the calculated $d_{x^2-y^2}$ cross-section for elastic processes, as shown in Fig. S3 b). This therefore suggests that the new Q ≈ 0.096 r.l.u. $d_{z^2}$ order exists on top of a standard scattering background coming from disordered $d_{x^2-y^2}$ orbitals in the bulk.

In Fig. S3 c) and d) we present additional calculations for scattering in $d_{z^2}$ orbitals. In c) we show the elastic counterpart of the spin flip process shown in the main text, while in d) we present an additional spin flip process with spins oriented along the [110] direction. Our current results do not allow us to differentiate between these different $d_{z^2}$ processes, but clearly imply ordering of the $d_{z^2}$ orbitals.

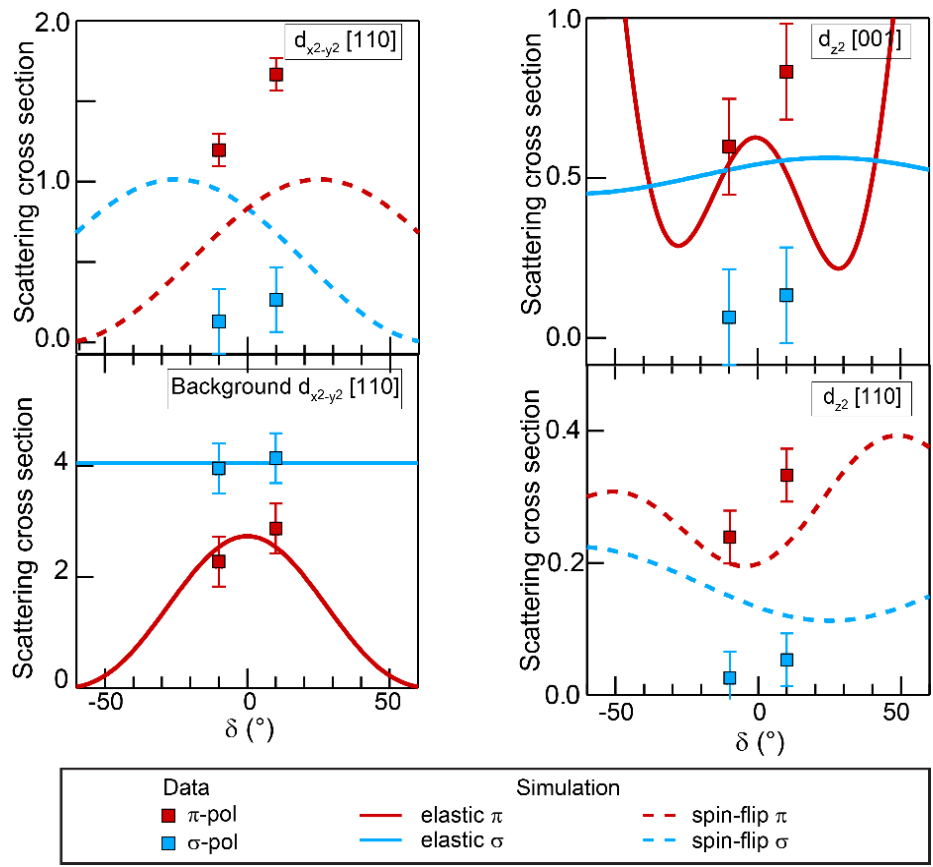

**Supplementary Figure 3. (a)** Calculated cross-sections for spin-flip scattering involving holes in the $d_{x^2-y^2}$ orbital, compared with the intensity enhancement (Bragg peaks) discussed in the main text. **(b)** Calculated cross-sections for elastic scattering involving holes in the $d_{x^2-y^2}$ orbital, compared with the measured background under the $Q_{//} \sim 0.1$ r.l.u. Bragg peak (symbols, ×4.5). The background data are consistent with a scattering process with $d_{x^2-y^2}$ character, implying a disordered background of has $d_{x^2-y^2}$ processes in the bulk. **(c) and (d)** Comparison of the measured cross sections of the $Q_{//} \sim 0.096$ r.l.u. Bragg peak (symbols, ×2 magnified) with calculated ones assuming **(c)** elastic scattering involving holes in the $d_{z^2}$ orbital, and **(d)** spin-flip scattering involving holes in the $d_{z^2}$ orbital for spins oriented along the [110] direction.

## Supplementary Note 4: REXS measurements of NYN trilayers with x=0.35

The Bragg peaks due to the Cu density wave orders described in the manuscript have also been observed with Resonant Elastic X-rays Scattering (REXS) at the UE46_PGM-1 beamline of the Bessy II synchrotron at the Helmholtz-Zentrum Berlin (HZB).[2,3] In the REXS experiment the energy of the diffracted x-ray photons is not analyzed, the Bragg peak of the elastic signal is therefore superimposed on a background that includes the entire inelastic signal and thus is much larger than in the RIXS experiment. The incident photon energy was set to the Cu-L$_3$ edge and tuned to the maximum of the intensity of the XAS signal, i.e. to 931.0 eV (in plane, σ-polarization) for the one with x=0.35 (see Fig. 4(c) of Ref.1). The scattering geometry of the REXS experiment was adjusted to probe the Bragg peak in the same configuration as in the RIXS experiments, i.e. at (H K L) = (-0.34 0 1.45) for x=0.35. In the REXS experiment, the value of H was scanned while keeping K and L constant by rotating the sample as well as the detector to vary the angle of incidence on the sample θ and the scattering angle 2θ.

Regarding the sample with x=0.35 and y=0.2, Supplementary Figure 4 demonstrates the presence of a Cu-CDW peak at H= -0.33 r.l.u and the absence of another Bragg peak, in particular around H= -0.096 r.l.u. This is evident from panel (b) which has been obtained by subtracting the curve taken at room temperature from the one at 6K.

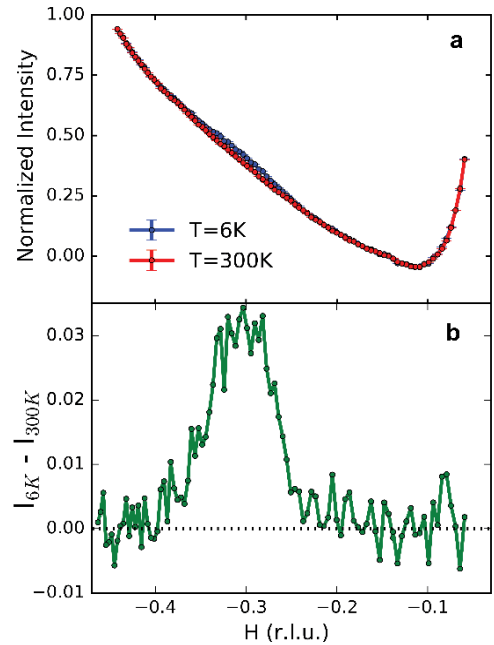

**Supplementary Figure 4.** *REXS study of a NYN trilayer with x=0.35, y=0.2 at 6K and 300K. **(a)** REXS measurement of the H scans at (-H 0 1.45) **(b)** The difference between the curves taken at 6K and 300K showing the charge order peak at H= -0.33 r.l.u and confirming the absence of a peak at H= -0.096 r.l.u..*

**Supplementary Note 5: Fitting the XAS data**

A quantitative analysis of the x-ray absorption spectroscopy (XAS) data in the region around the Cu-$L_3$ edge has been performed with a multicomponent peak fitting. The spectral weight of the absorption spectra has been normalized using the polarization averaged absorption curve, $(2\mu_{ab} + \mu_c)/3$. The resonant signal has been fitted with up to four Lorentzian functions for the trilayer with x=0.35 while only three Lorentzian functions were required to fit the data of the NYN trilayer with x=0.5. To minimize the number of free parameters, we used for each of the Lorentzian functions the same peak energy to fit simultaneously all the FY and TEY curves of a given sample.